\documentclass[aps,prd,preprint,groupedaddress,showpacs]{revtex4-1}
\usepackage{amsmath,amssymb}
\usepackage{graphicx}
\begin{document}
\title{Grand Rip and Grand Bang/Crunch cosmological singularities}
\author{L. Fern\'andez-Jambrina}
\email[]{leonardo.fernandez@upm.es}
\homepage[]{http://dcain.etsin.upm.es/ilfj.htm}
\affiliation{Matem\'atica Aplicada, E.T.S.I. Navales, Universidad
Polit\'ecnica de Madrid,\\
Arco de la Victoria 4, \\ E-28040 Madrid, Spain}%
\date{\today}
\begin{abstract}
The present accelerated expansion of the universe has enriched the
list of possible scenarios for its fate, singular or not.  In this
paper a unifying framework for analyzing such behaviors is proposed,
based on generalized power and asymptotic expansions of the barotropic
index $w$, or equivalently of the deceleration parameter $q$, in terms of the time coordinate.  Besides well known
singular and non-singular future behaviors, other types of strong
singularities appear around the phantom divide in flat models, with
features similar to those of big rip or big bang/crunch, which we have
dubbed \emph{grand rip} and \emph{grand bang/crunch} respectively,
since energy density and pressure diverge faster than $t^{-2}$ in
coordinate time.  In addition to this, the scale factor does not
admit convergent generalized power series around these singularities with a finite 
number of terms with negative powers.
\end{abstract}
\pacs{04.20.Dw, 98.80.Jk}

\maketitle

\section{Introduction}

Our universe is expanding acceleratedly, as it has been tested with
many sources of observational data \cite{snpion,
Davis:2007na,WoodVasey:2007jb,Leibundgut:2004,wmap}, and this fact has
led to several attempts of explanation, either by postulating the
existence of a new component of the energy of the universe, dubbed as
dark energy \cite{Padmanabhan:2006ag,Albrecht:2006um,Sahni:2006pa}, 
or by suggesting modifications of the theory of gravitation which 
would be consistent with observations at the cosmological scale 
\cite{Maartens:2007,Durrer:2007re,Padmanabhan:2007xy,modigravi}.

This has led to previously unregarded scenarios for the future 
behavior of the universe, since dark energy violates some of the 
conditions that were taken for granted for standard matter, such as 
the energy conditions \cite{HE}.

When all energy conditions were taken into account, the future 
evolution of our universe was restricted to collapse in a Big Crunch 
if the energy content were over a critical value or expansion forever 
if it were below such value.

Violation of energy conditions has increased the number of possible
singular fates from just Big Crunch to a list of new scenarios.  One
of the attempts to classify them \cite{Nojiri:2005sx} resorts to the
behavior of the scale factor $a(t)$, the Hubble ratio $H(t)$ and the
energy and pressure of the content of the universe at a value of time.
This classification has been refined and enlarged in \cite{IV}, 
\cite{yurov} and \cite{sesto}.  The latter one also includes other types of future
behaviors such as little rip and pseudo-rip.

\begin{itemize}    
   \item Type 0: ``Big Crunch'': Vanishing scale factor; blow up of Hubble ratio, energy density and 
   pressure.
   
   \item Type I: ``Big Rip'' \cite{Caldwell:2003vq}: Blow up of scale 
   factor, density and pressure. This was the first non-classical scenario 
   that was considered. Just the timelike geodesics are incomplete at 
   the Big Rip \cite{puiseux}, but not the lightlike ones.

   \item Type II: ``Sudden singularities'' \cite{sudden} (the first 
   model was introduced in \cite{suddenfirst}. They have been named
   ``quiescent singularities'' in the context of braneworld models 
   \cite{quiescent}): Finite scale
   factor, Hubble ratio and density; blow up of pressure.  They
   enclose Big Brake \cite{brake} and Big Boost \cite{boost} as a
   subcase.  These singularities do not violate the weak and strong
   energy conditions.  They are weak singularities \cite{suddenferlaz}
   and in this sense the universe can be extended after the singular
   event.

   \item Type III: ``Big Freeze'' \cite{freeze} or ``Finite Scale Factor 
   singularities'': Finite scale factor; blow up of Hubble ratio, 
   density and pressure. These can be either weak or strong 
   singularities depending on the criterion \cite{puiseux}.

   \item Type IV \cite{tsagas}: Finite scale factor, Hubble
   ratio, energy density and pressure; blow up of higher
   derivatives. These are also weak singularities. They are named 
   \cite{sesto} ``generalized sudden singularities'' if the barotropic index $w$ 
   remains finite, and Big Separation if it becomes infinite with 
   vanishing pressure and energy density.
   
   \item Type V: ``$w$-singularities'' \cite{wsing}, introduced in 
   the context of loitering braneworlds \cite{loitering}: Finite scale 
   factor, vanishing density and pressure. Just the barotropic index 
   $w$ blows up. These singularities are weak \cite{barotrope}.
\end{itemize}

Some of these singularities can be solved in loop quantum cosmology 
\cite{vidotto}.

To this list we could add another exotic type of singularities, which 
do not take place at a finite coordinate time:

\begin{itemize}

    \item Type $\infty$: ``Directional singularities'': These singularities are located 
    at infinite coordinate time, but some observers meet them in 
    finite normal time. Curvature scalars vanish there 
    \cite{hidden}, though they are strong singularities. 
\end{itemize}

Besides singularities, there are other types of future behavior, 
which are not singular, but mimic some of their features, though at 
an infinite time. They all have in common that for them the 
barotropic index is close to $w=-1$ and so they can be viewed as 
 deviations from $\Lambda$-Cold Dark Matter model:

\begin{itemize}
\item  Little Rip: \cite{little} At infinite time the Hubble ratio 
diverges. It shares the  features of Big Rip, but at an infinite time.

\item  Pseudo-rip: \cite{pseudo} Monotonic increase of the Hubble 
ratio, though finite, even at infinite time.

\item  Little Sibling of the Big Rip: \cite{sibling} The same 
features of Little Rip (the Hubble ratio blows up at infinite time), but with finite derivative of the Hubble 
ratio.
\end{itemize}

It would be interesting to unify all previous future behaviors,
singular or not, in one single framework or classification.  Our
proposal is to look at generalized power and asymptotic expansions in
coordinate time of the barotropic index $w$ (or the deceleration
parameter $q$).  We shall see that all future behaviors arise
naturally in this framework.  As a byproduct, new types of strong
singularities come up in the vicinity of the phantom divide $w=-1$,
sharing features of Big Crunch or Big Rip singularities, depending on
a sign, and so we have dubbed them \emph{grand crunch} and \emph{grand
rip} respectively, since energy density and pressure diverge faster
than $t^{-2}$ in coordinate time.  They have been overlooked in
previous frameworks since for them the scale factor does not admit
convergent generalized power expansions around the singularity with a
finite number of terms with negative powers, though the barotropic
index, the energy density and the pressure do.  We shall focus on
them.

The paper is organizes as follows.  In Section~\ref{w} we solve the
Friedman equations for a FLRW cosmological model in terms of the
barotropic index $w$.  This is shown useful to postulate several kinds
of behavior for $w$, such as power expansions at at finite time event
or asymptotic expansions at infinity, which we deal with in
Section~\ref{infinito}, and translate them to the scale factor, the
energy density and pressure of the universe.  Features of the new
types of singularities as well as their geodesic incompleteness and
strength are analyzed respectively in Sections~\ref{great} and
\ref{strength}.  We end up with a section of Conclusions.

\section{Singularities, barotropic index $w$ and deceleration parameter $q$\label{w}}
    
We  consider spatially flat homogeneous and isotropic spacetimes with 
a metric tensor of the form
\begin{equation}ds^2=-dt^2+a^2(t)\left(dr^2+ r^2\left(d\theta^2+\sin^2\theta
d\phi^2\right)\right),\label{metric}\end{equation}
where $a(t)$ is the scale factor of the universe in cosmological time 
$t$. Einstein equations for such spacetimes reduce to the usual Friedman 
equations,
\begin{equation}\label{flrw}\rho=
\frac{3\dot a^2}{a^2},\qquad 
p=-\frac{2\ddot a}{a}-\frac{\dot a^2}{a^2},\end{equation}
in terms of the energy density $\rho(t)$ and pressure $p(t)$ of 
the content of the universe. The dot stands for derivation with 
respect to the time coordinate. We are using geometrized units for 
which $c=1=8\pi G$.

Defining the time-dependent barotropic index of the 
universe $w(t)$ as the ratio between pressure and energy density 
allows us to write it in terms of the scale factor and its derivatives,
\[
w=\frac{p}{\rho}=-\frac{1}{3}-\frac{2}{3}\frac{a\ddot a}{\dot a^2}.\]

This formula is valid just for flat models. If curvature is taken 
into account, additional terms are necessary.

Tha barotropic index $w$ is closely related to the deceleration 
parameter $q$,
\[q=-\frac{a\ddot a}{\dot a^2}=\frac{1+3w}{2},\]
again for flat models. Otherwise the relation between both parameters 
becomes more complicated, involving also the Hubble parameter $H=\dot 
a/a$. This allows direct translation of our results for the 
barotropic index to the deceleration parameter.

We may see this equation the other way round as the differential 
equation governing the evolution of the scale factor for a given 
barotropic index $w(t)$. In fact, we may appease its non-linearity by 
introducing the time function $x=\ln a$,
\[\frac{\ddot x}{\dot x^2}=-\frac{3}{2}(w+1)=-(q+1),\]
which suggests defining
\[h(t):=\frac{3}{2}(w(t)+1)=q(t)+1\] as a correction around the pure 
cosmological constant case,
\[ w(t)=-1+\frac{2}{3}h(t),\qquad q(t)=-1+h(t).\]

This change of variables helps us lower the order of the 
differential equation,
\[h=-\frac{\ddot x}{\dot x^2}=\left(\frac{1}{\dot 
x}\right)^\cdot \Rightarrow \dot x=\left(\int h\,dt 
+K_{1}\right)^{-1},\]
which can be solved as a quadrature in terms of two free constants $K_{1}$, 
$K_{2}$,
\[a(t)=\exp\left(\int\left(\int h(t)\,dt 
+K_{1}\right)^{-1}dt+K_{2}\right).\]

The constant $K_{2}$ is part of a global constant factor 
$a(t_{0})=\exp(K_{2})$,
\begin{equation}\label{scale}
a(t)=a(t_{0})\exp\left(\int_{t_{0}}^t\left(\int h(t)\,dt 
+K_{1}\right)^{-1}dt\right),\end{equation}
which is fixed by the choice of scale factor equal to one nowadays. 
Models with such exponential behavior can be found in 
\cite{IV}.

For fixing $K_{1}$ we are to resort to one of the Friedman equations 
(\ref{flrw}), since we have made use of just the ratio between 
pressure and energy density,
\[\rho(t)=3\dot x(t)^2=3\left(\int_{t_{0}}^t h(t)\,dt 
+K_{1}\right)^{-2},\]
\[p(t)=-2\ddot x(t)-3\dot x(t)^2=\frac{2h(t)-3}{\left(
\displaystyle\int_{t_{0}}^t h(t)\,dt 
+K_{1}\right)^{2}},\]
allowing us to determine $K_{1}=\sqrt{3}\rho(t_{0})^{-1/2}$, unless $\rho$ 
becomes infinite at $t=t_{0}$, in which case $K_{1}=0$.

We focus on the latter case since our interest is the possibility of 
formation of singularities. In order to simplify the 
notation, a time translation is performed to locate the singular 
event at $t=0$. The global factor due to $K_{2}$ is also omitted.

From the expression for the scale factor,
\[a(t)=\exp\left(\int\frac{dt}{\int h(t)\,dt}\right),\]
we learn that there are several qualitative behaviors depending on 
the rate of growth of $h(t)$. If we assume that this function can be 
expanded in powers of time around $t=0$,
\[h(t)=h_{0}t^{\eta_{0}}+ h_{1}t^{\eta_{1}}+ \cdots, \qquad 
\eta_{0}<\eta_1<\cdots,\]
we get expressions for the  scale factor, the energy density and the 
pressure at lowest order in $t$,
\[x(t)=\left\{\begin{array}{ll}\displaystyle-\frac{\eta_{0}+1}{\eta_{0} 
h_{0}}t^{-\eta_{0}}+\cdots &\textrm{if\ } -1\neq \eta_{0}\neq 0\\\\
\displaystyle \frac{1}{h_{0}}\int\frac{dt}{\ln|t|}+\cdots &\textrm{if\ } 
\eta_{0}=-1\\\\
\displaystyle\frac{\ln|t|}{h_{0}}+\cdots
&\textrm{if\ }\eta_{0}=0.
\end{array}\right.\]

For simplicity, we have considered $t>0$. Since our equations are 
symmetric under time reversal, the same  expressions are valid 
exchanging $t$ by $-t$ in order to consider times before $t=0$.

Once we know the scale factor, we can derive expressions at lowest
order for the energy density,
\[\rho(t)=\left\{\begin{array}{ll}\displaystyle 3\left(\frac{\eta_{0}+1}{h_{0}}\right)^2t^{-2(\eta_{0}+1)}+\cdots
&\textrm{if\ }-1\neq \eta_{0}\neq 0\\\\\displaystyle 
\frac{3}{ h_{0}^2}\frac{1}{\ln^{2}|t|}+\cdots&\textrm{if\ }\eta_{0}=-1
\\\\\displaystyle
\frac{3t^{-2}}{h_{0}^2}
+\cdots 
&\textrm{if\ }\eta_{0}= 0,\end{array}\right.\]
and the pressure,
\[p(t)=\left\{\begin{array}{ll}\displaystyle
\frac{2(\eta_{0}+1)^2}{h_{0}}t^{-\eta_{0}-2}+\cdots 
&\textrm{if\ }-1\neq\eta_{0}<0\\\\\displaystyle 
\frac{2}{ h_{0}}\frac{1}{t\ln^{2}|t|}+\cdots &\textrm{if\ 
}\eta_{0}=-1\\ \\\displaystyle
\frac{2h_{0}-3}{h_{0}^2}t^{-2}
+\cdots  &
\textrm{if\ }\eta_{0}=0\\
\\ -\displaystyle
3\left(\frac{\eta_{0}+1}{h_{0}}\right)^2t^{-2(\eta_{0}+1)}+\cdots &
\textrm{if\ }\eta_{0}>0,
\end{array}\right.\]
and we come across several possibilities:
\begin{itemize}
    \item For $\eta_{0}<-2$, both $\rho$ and $p$ vanish at $t=0$
    whereas $w$ diverges.  These are generalized sudden or type IV 
    singularities. They also comprise the kind of singularities
    discussed in \cite{wsing} and \cite{barotrope}, which include
    $w$-singularities, for which all derivatives of the energy density
    and pressure are regular, but with just diverging barotropic index.

    \item For $\eta_{0}=-2$, $\rho$ vanishes at $t=0$ as $t^2$, but 
    $p$ remains finite,  whereas $w$ diverges. They are a special 
    case of generalized sudden singularities.

    \item  For $\eta_{0}\in(-2,-1]$, $\rho$ vanishes at $t=0$, but 
    $p$ diverges. These are sudden or type II singularities \cite{sudden}.

    \item  For $\eta_{0}\in (-1,0)$, $\rho$, $p$ and $w$ diverge at 
    $t=0$. These are type III, Big Freeze of finite scale factor 
    singularities.
    
    \item For $\eta_{0}=0$, both $\rho$ and $p$ diverge at $t=0$ as 
    $t^{-2}$ and  $w\simeq -1+2h_{0}/3$ is finite, corresponding to 
    models of the form $a(t)\simeq t^{1/h_{0}}$. These produce 
    classical Big Bang/Big Crunch singularities if $h_{0}$ is 
    positive and Big Rip or type I singularities if $h_{0}$ is negative.
    
    \item For $\eta_{0}>0$, $\rho$ and $p$ diverge at $t=0$ as
    $t^{-2(\eta_{0}+1)}$ and $w$ tends to the value $-1$.  The
    possibility of singularity has not been considered before in the
    previous frameworks.  The reason for this is that it cannot be
    embedded in the classifications in \cite{puiseux} and
    \cite{visser}, since the scale factor
    (exponential of rational functions)  does not accept
    convergent power expansions, generalized or not, with a finite 
    number of terms with negative powers, though $x(t)$
    does.  We name them \emph{grand rip} or \emph{grand bang/crunch},
    depending on the behavior of the scale factor at the singularity.
    We analyze these in detail in Section~\ref{great}.
\end{itemize}

There is a case of singular pressure with finite energy density when 
$K_{1}\neq 0$. To achieve this we need $h(t)$ diverging at $t_{0}$, 
that is, $\eta_{0}<0$, but the integral of $h(t)$ must be finite, 
which implies $\eta_{0}>-1$. Hence, we have finite energy density and 
infinite pressure at $t_{0}$ if $\eta_{0}\in(-1,0)$. This case 
corresponds to a sudden or type II singularity with finite energy 
density.

These results are summarized in Table~\ref{tablw}, where we have 
related the first exponent in the generalized power expansion of 
$h(t)$ at the singularity to the values of the scale factor, the energy density, the 
pressure and the barotropic index and to the type of singularity.

\begin{table}[h]
   \begin{tabular}{cccccccc}
   \hline
   ${\eta_{0}}$ &$a_{s}$ & $\rho_{s}$ &$p_{s}$ & $w_{s}$ & Sing.\\
   \hline
   $(-\infty,-2)$ &  finite &   0 & 0 & $\infty$ & IV or V\\ 
   $-2$ & finite &    0 & finite & $\infty$ & IV \\
   $(-2,-1]$ & finite &  0 &  $\infty$ & $\infty$ & II  \\
   $(-1,0)$, $K_{1}\neq0$  & finite 
      & finite & $\infty$ & $\infty$ &  II 
      \\
   $(-1,0)$, $K_{1}=0$ & finite 
      & $\infty$ & $\infty$ & $\infty$ & III \\
0 &  0/$\infty$ & $\infty$ & $\infty$ & finite & big crunch / rip \\
   $(0,\infty)$ &0/$\infty$&   $\infty$ & $\infty$ & -1 & grand 
   crunch / rip \\
   \hline
   \end{tabular}
\caption{Expansions of $q$ and $w$ at $t_{s}$ vs. possible singularities}\label{tablw}
\end{table}

\section{Behavior at infinite time\label{infinito}}

In addition to this analysis of singularities at a finite coordinate 
time $t$, we can take into account what happens at $t=\infty$. It is 
not pointless, since it has been shown \cite{hidden} that there are 
geodesics in FLRW spacetimes which reach $t=\infty$ in a finite 
proper time. As we have already pointed out, the analysis for 
$t=-\infty$ is entirely similar.

For this analysis we consider now asymptotic expressions for $h(t)$ 
for large $t$. We take then $t_{0}=\infty$ in (\ref{scale}). Asymptotic expressions for the 
scale factor, the energy density and pressure take the form
\[
a(t)=\exp\left(-\int\left(\int_{t}^\infty h(t)\,dt+K_{1}\right)^{-1}dt\right),\]
\[\rho(t)=3\left(\int_{t}^\infty h(t)\,dt +K_{1}\right)^{-2},\]
\[p(t)=\frac{2h(t)-3}{\left(
\displaystyle\int_{t}^\infty h(t)\,dt+K_{1} \right)^{2}}\]
and if the constant $K_{1}=0$,  $\rho$ and $p$ diverge at infinity. 

Of course, these expressions are valid only if the integral 
\begin{equation}\label{intinf}
\int_{t}^{\infty}h(t)\,dt\end{equation} is finite.  With this we
guarantee that $K_{1}=\sqrt{3}\rho(\infty)^{-1/2}$, which is useful
for keeping control of the asymptotic behavior of the energy density.
Otherwise, we would have to resort to expressions (\ref{scale}) for
large $t$.

For having a finite integral (\ref{intinf}) we need 
$h(t)\to 0$ for large $t$, though it is not a sufficient condition. 
For instance, $h(t)=1/t$ tends to zero, but its integral diverges for 
large $t$. Combining finiteness of (\ref{intinf}) and asymptotic 
behavior of $h(t)$  leads in principle to several cases:

\begin{itemize}
    \item  Finite $\displaystyle\int_{t}^{\infty}h(t)\,dt$: This 
happens when $h(t)$ decreases faster than $1/t$. 
We consider first this case.

Since $h(t)$ tends to zero for large values of time, the asymptotic 
value of the barotropic index $w$ is -1:

\begin{itemize}
    \item  If $h(t)>0$ for large values of $t$, the scale factor 
    decreases to zero at infinity as a negative exponential. It 
    would be a sort of \emph{little crunch}.  The asymptotic value 
    $w_{\infty}=-1$ of the barotropic index is reached from above in 
    this case. Since $a(t)$ is an integrable function at infinity, 
    this case is included in the set of directional singularities 
    described in \cite{hidden}, which are strong singularities, but 
    only accessible for some observers.

    \item  If $h(t)<0$ for large values of $t$, the scale factor 
    blows up at infinity exponentiallly. It 
    is the Little Rip \cite{little} or, for some choices of $h(t)$, 
    the Little Sibling \cite{sibling}.  The asymptotic value 
    $w_{\infty}=-1$ of the barotropic index is reached from below.
\end{itemize}

If we let $K_{1}\neq 0$, the scale factor, the energy density and the 
pressure would be finite at infinity. The case $K_{1}<0$ would correspond to a Pseudo-rip \cite{pseudo}.
%
%
%

    \item  Infinite $\displaystyle\int_{t}^{\infty}h(t)\,dt$: The expression 
for the scale factor (\ref{scale}), as well as the ones for the 
     energy density and the pressure are valid with $K_{1}\neq0$. In 
    this case both the energy density and the pressure tend to zero 
    for large $t$. The sign of $h(t)$, as in the previous case, 
    determines if the scale factor diverges or tends to zero. The asymptotic value of the barotropic index 
    $w_{\infty}$ is -1 if $h(t)$ tends also to zero. This leads to 
    several subcases:
    \begin{itemize}
    \item  $1/t\lesssim |h(t)|\to 0$ for large $t$: The asymptotic 
    value of the scale factor is $w_{\infty}=-1$. 
    
    If $h(t)$ is 
    negative for large $t$, the scale factor decreases exponentially 
    as an integrable function. This means that non-comoving observers 
   and lightlike geodesics \cite{hidden} take finite normal time to reach 
    time to reach $t=\infty$, which is a strong \emph{directional 
    singularity}.
    
    If $h(t)$ is positive for large $t$, the scale factor increases
    exponentially and so this case is similar to the \emph{little 
    rip}, but with asymptotically vanishing energy density and 
    pressure and approaching the asymptotic value $w_{\infty}=-1$ from 
    above.
    
    \item  $h(t)\sim K$ const. for large $t$: The asymptotic value of 
    the scale factor is $w_{\infty}=-1+2K/3$ and the scale factor 
    behaves as a power of time, $a(t)\sim t^{1/K}$, which is an 
    integrable function for large $t$ if $K\in(-1,0)$, 
    corresponding to a strong \emph{directional singularity} at 
    $t=\infty$. Otherwise, the scale factor diverges for $K>0$ or 
    tends to zero for $K\le-1$, but without singularity.
    
    \item $|h(t)|\to\infty$ for large $t$: The barotropic index 
    diverges and the scale factor is non-integrable. There is no 
    singularity in this case. 
    
    If $h(t)$ is positive for large $t$, the scale factor grows to a
    finite asymptotic constant value if $\int dt/\int h(t)\,dt$
    converges.  Otherwise, the scale factor diverges to infinity. 
    
    If $h(t)$ is negative for large $t$, the scale factor decreases 
    to a finite asymptotic constant value if $\int dt/\int h(t)\,dt$ converges. 
    Otherwise, the scale factor tends to zero.
    \end{itemize}
\end{itemize}

These results are summarized in Table~\ref{tablinf}, where the 
asymptotic behavior of $h(t)$ for large $t$ is related to the 
asymptotic values of the scale factor, the energy density, the 
pressure and the barotropic index and to the type of singularity or 
future behavior.

\begin{table}[h]
   \begin{tabular}{ccccccccc}
   \hline
   $h$ & signum\,($h$) & $K_{1}$ &$a_{\infty}$ & $\rho_{\infty}$ &$p_{\infty}$ & $w_{\infty}$ & 
   Behavior\\
   \hline Finite
$\int^\infty h\,dt$ & + & 0& 0 &   $\infty$ &  $\infty$ & -1 &  $\infty$
\\ & - & 0& $\infty$ &   $\infty$ &  $\infty$ & -1 &  little rip / sibling
\\  & $\pm$ & positive & 0 &   finite &  finite & -1 &  non-singular
\\  & $\pm$ & negative & $\infty$ &   finite &  finite & -1 &  pseudo-rip
\\ $t^{-1}\lesssim |h(t)|\to 0$ & +& any  & $\infty$ &   0 &  0 & -1 & little 
rip with 0 $\rho$ and $p$ 
\\  & - & any& 0 &   0 &  0 & -1 &  $\infty$ 
\\ $K$ & +& any & $\infty$ &   0 &  0 & -1+2K/3 &  non-singular
\\ $K\in (-1,0)$ &- & any &  0 &   0 &  0 & -1+2K/3 &  $\infty$
\\$K\in  (-\infty,-1]$ & -& any  & 0 &   0 &  0 & -1+2K/3 & non-singular 
\\   $|h(t)|\to \infty$, \ infinite $\int^\infty dt/\int h(t)\,dt$& + & any& 
$\infty$ &   0 &  0 & $\infty$ & non-singular 
\\   $|h(t)|\to \infty$, \ infinite $\int^\infty dt/\int h(t)\,dt$& - 
& any& 0 &   0 &  0 & $\infty$ & non-singular 
\\   $|h(t)|\to \infty$, \ finite $\int^\infty dt/\int h(t)\,dt$& $\pm$ & any& finite &   0 &  0 & $\infty$ & non-singular 
\\  
\hline
   \end{tabular}
\caption{Asymptotic behavior of $q$ and $w$ at $t=\infty$ vs. possible 
behaviors}\label{tablinf}
\end{table}

\section{Grand rip and grand bang/crunch singularities\label{great}}
    
Let us take a look at the new family of singularities for $\eta_{0}>0$. First of all, 
we notice that pressure and energy density diverge as a power of 
coordinate time which is different from -2, which would be the case of Big 
Bang/Crunch and Big Rip, but it can be as close to such value  as 
desired if the exponent $\eta_{0}$ is small enough.

Second, whereas Big Bang/Crunch and Big Rip have a different value of
the barotropic index $w(0)$ depending on the equation of state, these
singularities have the value $w(0)=-1$ regardless of the exponent
$\eta_{0}$.  Considering only the barotropic index, these
singularities arise as small perturbations in coordinate time,
$w(t)=-1+2h_{0}t^{\eta_{0}}/3$, $\eta_{0}>0$, around the de Sitter value.
This does not mean of course that such perturbations are necessarily
singular, since we have explicitly removed the constant $K_{1}$ in
order to look for singular behavior.

The sign of the coefficient $h_{0}$ determines the type of 
singularity. Since 
\[a(t)\simeq e^{-\mathrm{sgn}\,(h_{0})\alpha/t^{\eta_{0}}}, \qquad
\alpha=\frac{\eta_{0}+1}{\eta_{0}|h_{0}|}>0,\qquad t>0,\]
we observe two kinds of behavior:

\begin{itemize}
\item $h_{0}>0$: In this case the exponential in (\ref{scale})
decreases for $t>0$ and the scale factor $a$ tends to zero on
approaching $t=0$ (Figure~1 left).  This would be a sort of exponential Big Bang
singularity, or Big Crunch if we exchange $t$ for $-t$.  Since $h_{0}$
is positive, the barotropic index $w$ remains always under the phantom
divide close to $t=0$.  That is, the value $w=-1$ is approached from
below. In order to pinpoint the differences and similarities with 
classical Big Bang and Big Crunch singularities we may call them 
\emph{grand bang} and \emph{grand crunch} singularities.

\item $h_{0}<0$: On the contrary, the exponential increases for $t>0$,
and the scale factor $a$ diverges to infinity on approaching $t=0$ 
(Figure~1 right). We would have then a sort of exponential Big Rip at $t=0$, which we
can locate in the future by exchanging $t$ for $-t$.  In this case the
barotropic index $w$ is always over the phantom divide and hence the
value $w=-1$ is approached from above. As in the previous case, we 
may name them \emph{grand rip} singularities.
\end{itemize}
\begin{figure}
\includegraphics[height=0.3\textheight]{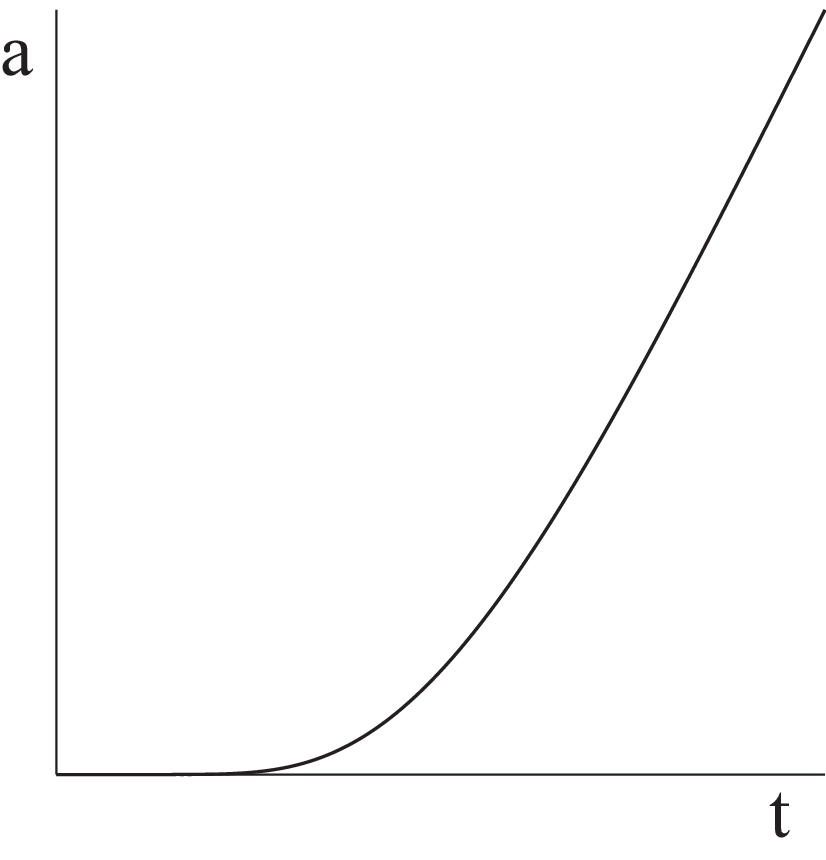}
\includegraphics[height=0.3\textheight]{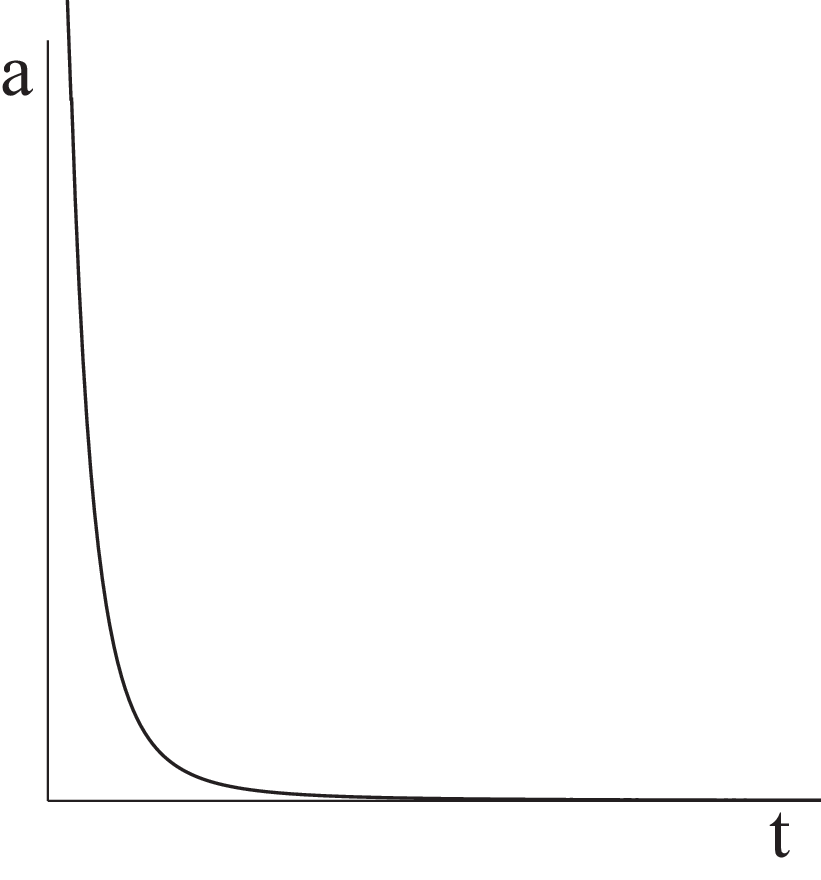}
\caption{Singularities at $t=0$ for $\eta_{0}=1$}
\end{figure}

We may check the behavior of causal geodesics at these singularities 
\cite{HE}. 
We consider parametrized curves on a FLRW spacetime, $\gamma(\tau)=(t(\tau), 
r(\tau), \theta(\tau), \phi(\tau))$, and impose a normalization 
condition on the velocity $u(\tau)=\gamma'(\tau)$, depending on its causal type
\begin{equation}
\left.\begin{array}{lc}\textrm{Timelike:}& -1\\\textrm{Lightlike:} &0\\ 
\textrm{Spacelike:} &+1 \end{array}
\right\}=\varepsilon=\|\gamma'(\tau)\|^2=-t'^2(\tau)+a^2(t(\tau))
( r'^2(\tau)+r^2(\tau)(\theta'^2(\tau)+\sin^2\theta(\tau) \phi'^2(\tau)),
\label{geodesic}\end{equation}
where the prime denotes derivation with respect to the parameter 
$\tau$.

Geodesic curves have zero acceleration, $\nabla_u u\equiv 0$, where
$\nabla$ is the covariant derivative associated to the metric
(\ref{metric}).  However, in this simple case, there is no need to
write down the whole system of second order differential equations
\cite{puiseux}.  Taking into account the symmetry of FLRW spacetimes,
it suffices for our analysis to consider curves on the equatorial
hypersurface $\theta=\pi/2$ with constant angle $\phi$. Homogeneity 
of the spacetime implies that the linear momentum of geodesics is a 
conserved quantity,
\[P=u\cdot \partial_{r}=a^2{(t)}r'.\]

This equation together with the normalization condition
(\ref{geodesic}) allow us to write the set of differential
equations for geodesic motion as a first order system
\[ r'=\frac{P}{a^2(t)},\qquad t'=\sqrt{-\varepsilon+\frac{P^2}{a^2(t)}},
\] for the normal parameter $\tau$.

The key equation is the second one, since the first equation can be
integrated once $t(\tau)$ is known.

We analyze know whether causal geodesics are complete \cite{HE}, that is, if 
the parameter $\tau$ can be extended from $-\infty$ to $\infty$. 

It this happens, it would take an infinite normal time to reach the 
singularity, which would not be accessible along causal 
geodesics. This would mean that the wordlines of non-accelerated 
observers traveling along them would not end up there.

The analysis of causal geodesics in FLRW spacetimes reduces to just three families of 
curves:

\begin{itemize}
\item  Lightlike geodesics: $\varepsilon=0$. These are readily solved,
\[t'= \frac{P}{a(t)} \Rightarrow 
\tau=\frac{1}{P}\int_{0}^{t}a(t)\,dt,\]if the scale factor is 
integrable.

In our case, $a(t)\simeq 
e^{-\mathrm{sgn}\,(h_{0})\alpha/t^{\eta_{0}}}$, the integral is 
convergent for positive $h_{0}$. This means that lightlike geodesics meet 
the singularity at $t=0$ in a finite normal time $\tau$. These 
geodesics are therefore incomplete.

On the contrary, for negative $h_{0}$, the integral is divergent and 
it takes an infinite normal time $\tau$ to reach $t=0$. Hence in this case 
lightlike geodesics avoid reaching the singularity and are complete 
in that direction. This is similar 
to what it happens for Big Rip singularities \cite{puiseux}.

\item Comoving timelike geodesics: $\varepsilon=-1$, $P=0$. In this 
case we can take $\tau=t$ and in both cases these geodesics meet the 
singularity. They are incomplete.

\item Radial timelike geodesics: $\varepsilon=-1$, $P\neq0$. 
For $h_{0}>0$, we have  $a(t)\ll 1$ and hence
\[t'=\sqrt{1+\frac{P^2}{a^2(t)}}\simeq \frac{P}{a(t)},\]
and so this case is similar to the lightlike one. They are incomplete.

For $h_{0}<0$, we have  $a(t)\gg 1$ and $t'\simeq 1$ and we arrive 
at the same conclusions as in the comoving case. They are also 
incomplete.
\end{itemize}

Summarizing, all causal geodesics arrive at $t=0$ in finite normal
time and are thereby incomplete except for lightlike geodesics at the 
\emph{grand rip}, which
are complete and do not experience the singularity.

\section{Strength of grand rip and grand crunch singularities\label{strength}}

Finally, we can check if the strength of the new singularities at $t=0$ is
enough for tidal forces to distort extended bodies \cite{ellis}.  There
are several criteria to determine this.  All of them model the finite
object at each point of a causal geodesic by a volume spanned by three
independent Jacobi fields in the hyperspace which has as normal vector
the velocity of the curve.  Tipler's criterion \cite{tipler} considers
that a singularity is strong if such volume tends to zero on
approaching the singularity along the geodesic, whereas Kr\'olak's
criterion \cite{krolak} just demands that the derivative of the
volumen with respect to the normal parameter must be negative.  Hence,
there are singularities which are strong according to Kr\'olak's
criterion, but  weak according to Tipler's,  for instance, type III
or Big Freeze singularities \cite{puiseux}. Another criterion can be 
found in \cite{rudnicki}.

Dealing with Jacobi fields is burdensome, since it involves solving
the Jacobi equation along geodesics.  However, characterizations for
lightlike geodesics and necessary and sufficient conditions for 
timelike geodesics for
fulfillment of both criteria have been established \cite{clarke} in
terms of integrals  of the Ricci and Riemann curvatures
of the metric of the spacetime along these curves:

\begin{itemize}
\item Lightlike geodesics: According to Tipler's criterion a
singularity is strong along a lightlike geodesic if and only if
\[\displaystyle
\int_{0}^{\tau}d\tau'\int_{0}^{\tau'}d\tau''R_{ij}u^{i}u^j\] 
blows up when the normal parameter $\tau$ approaches the value 
corresponding to the singularity.

According to Kr\'olak's criterion the singularity is strong
if and only if the integral
\[\displaystyle \int_{0}^{\tau}d\tau'R_{ij}u^{i}u^j,\]
blows up when $\tau$ approaches the singularity.

In our case, $u=(t',r',\theta',\phi')=(P/a,P/a^2,0,0)$, integrals of
\[R_{ij}u^iu^j\,d\tau=2P^2\left(\frac{a'^2}{a^4}-\frac{a''}{a^3}\right)\frac{a\,dt}{P}
\simeq 
\frac{2P\mathrm{sgn}\,(h_{0})\alpha\eta_{0}(\eta_{0}+1)}{t^{\eta_{0}+2}}
e^{\mathrm{sgn}\,(h_{0})\alpha/t^{\eta_{0}}}dt\]
blow up at $t=0$ for all $h_{0}>0$ and hence these singularities are 
strong according to both criteria. For $h_{0}<0$ we already know that 
these geodesics do not even reach the singularity.

\item  Timelike geodesics: For these curves \cite{clarke} 
does not provide a characterization, but different necessary and 
sufficient conditions. 

Following Tipler's criterion  a
singularity is strong along a timelike geodesic if
\[\displaystyle\int_{0}^{\tau}d\tau'\int_{0}^{\tau'}d\tau''R_{ij}u^{i}u^j\]
blows up on approaching the singularity.
 
Following Kr\'olak's criterion, the singularity is strong
if  the integral
\[\displaystyle\int_{0}^{\tau}d\tau'R_{ij}u^{i}u^j \]
blows up on approaching the singularity.

There are also necessary conditions, but we are not making use of 
them for our purposes.

For comoving geodesics, $u=(1,0,0,0)$, integrals of
\[R_{ij}u^i u^j\,d\tau=-\frac{3a''}{a}\,dt\simeq 
-\frac{3\alpha^2\eta_{0}^2}{t^{2\eta_{0}+2}}dt,\]
blow up for all $\eta^0>0$ and therefore singularities at $t=0$ are 
strong.

For radial geodesics, 
$u=\left(\sqrt{1+P^2/a^2},\pm P/a^2,0,0\right)$, the analysis is 
similar,
\[R_{ij}u^i u^j\,d\tau=
\frac{-\frac{3a''}{a}+2P^2\left(\frac{a'^2}{a^4}-\frac{a''}{a^3}
     \right)}{\sqrt{1+\frac{P^2}{a^2}}}dt\simeq dt
\left\{\begin{array}{cl}
-\frac{3a''}{P}+2P\left(\frac{a'^2}{a^3}-\frac{a''}{a^2} \right)
&\textrm{if }a\to 0\\
-\frac{3a''}{a}+2P^2\left(\frac{a'^2}{a^4}-\frac{a''}{a^3} \right)
&\textrm{if }a\to \infty.
\end{array}\right.\]

For $h_{0}>0$, $a,a''$ tend to zero at $t=0$, but the $P$ term has 
been shown to be exponentially divergent.

For $h_{0}<0$, the integrals of the $a''/a$ term have been shown to 
be divergent, though the $P$ term tends to zero.

Hence, in both cases radial geodesics meet a strong singularity at 
$t=0$.

Summarizing, singularities are strong for all geodesics except for 
lightlike geodesics in the $h_{0}<0$ case, which are not even 
incomplete.

\end{itemize}

\section{Concluding remarks}

We have shown that generalized power and asymptotic expansions of the
barotropic index $w$ and the deceleration parameter $q$ in time coordinate are useful to classify most
singular and non-singular future behaviors of the universe.  In
addition to well known scenarios, another type of possible singular
behavior is found for small corrections of $w=-1$ and $q=-1$ at a finite time.
These singularities share many features of big rip or big bang/crunch
singularities, depending on the sign of the perturbation, and so we
have dubbed them respectively \emph{grand rip} and \emph{grand bang/crunch}
singularities. They can appear just when the barotropic index and the 
deceleration parameter take the 
value of minus one.  Both energy density and pressure diverge at the
singularity as a negative power of coordinate time, which can be as
close as desired to minus two. The scale factor does not admit power 
expansions around the singular value $t=0$ with a finite 
number of terms with negative powers, not even in the case of 
vanishing $a(0)$. They are strong singularities,
except for lightlike geodesics, which avoid the \emph{grand rip} 
singularity.

Considering the asymptotic expansions at $t=\infty$, in addition to 
little rip and pseudo-rip behaviors, the only singularities
that are found are directional singularities, which are experienced 
just by non-comoving observers and lightlike geodesics. As a novelty, 
they are also found for $w_{\infty}=-1$.

\section*{Acknowledgments}
The author wishes to thank the referees for their useful comments.
    
\end{document}